\documentclass{article}

\usepackage{amsmath}
\usepackage{subfigure}
\usepackage{url}
\usepackage[utf8]{inputenc}

\usepackage[pdftex]{graphicx}

\usepackage{mathtools}
\DeclarePairedDelimiter{\ceil}{\lceil}{\rceil}
\usepackage{amsmath}

\usepackage[utf8]{inputenc}
\usepackage{epsfig}
\usepackage{epstopdf}
\usepackage{graphicx}
\usepackage{times}
\usepackage{url}
\usepackage{multirow}
\usepackage{algorithm}
\usepackage{amsmath,amsthm}
\usepackage{amsfonts,amssymb}
\usepackage{multirow}
\usepackage{subfigure}
\usepackage{tabularx}
\usepackage{comment}
\usepackage{xspace}
\usepackage{soul}
\usepackage{color}
\usepackage{booktabs}
\usepackage{diagbox}
\usepackage{algpseudocode}
\usepackage{enumerate}

\usepackage{authblk}
 
\author[1]{Francesca Cuomo}
\author[2]{Manuel Campo}
\author[1]{Enrico Bassetti}
\author[1]{Lorenzo Cartella}
\author[1]{Federica Sole}
\author[3]{Giuseppe Bianchi}
\affil[1]{University of Rome "La Sapienza", Italy}
\affil[2]{CNIT, Italy}
\affil[3]{University of Rome "Tor Vergata", Italy}

\date{}

\begin{document}
	\title{Adaptive mitigation of the Air-Time pressure in LoRa  multi-gateway architectures} 
 
\maketitle

\begin{abstract}
LoRa  is  a  promising  technology  in  the  current  Internet of Things
market, which operates in un-licensed bands achieving long-range
communications and with ultra power devices.
In this work we capitalize on the idea introduced in \cite{Explora-AT},
i.e. balance the Air-Time of the different modulation spreading factors
(SF), and adapt it to operate in a typical metropolitan scenario
comprising multiple gateways (GWs) interconnected to a same network
server. Our proposed approach, named ADaptive Mitigation of the AIr-time
pressure in lORa (AD MAIORA), relies on a suitable measure of the
per-spreading-factor load at each GW - quantified by means of a
so-called \textit{pressure table} -, and on a relevant heuristic
algorithm which attempts to balance such a \textit{per-SF-pressure}.
Especially in cases of very loaded scenarios, where a high number of
nodes insist on the same GWs, the use of AD MAIORA shows significant
performance gains, up to a factor of 5 improvements with respect to the
legacy LoRaWAN's Adaptive Data Rate.

\end{abstract}

\section{Introduction}
LoRa  is  a  promising  technology  in  the  current  Internet of Things (IoT) market. It operates in the ISM  band  using  a  proprietary  spread spectrum  technique developed  and  commercialized  by
Semtech  Corporation  \cite{LoRa}\cite{LoraforIoT}. It achieves long-range  communications and supports ultra low power devices.
The  LoRaWAN  network  comprises multiple gateways (GWs), operating in a wide-area, and providing connectivity to the possibly huge amount of deployed end-devices.
The LoRa modulation is based on the Chirp Spread Spectrum technique  offering low transmission power and robustness from channel degradations.  Transceivers in the LoRaWAN GW receive  multiple  number
of  messages  from  the  used channels  \cite{Alliance}. The  spread  spectrum  provides orthogonal separation between signals by using different spreading factors to the individual signal.
Thanks to this, LoRa provides a  bidirectional communication at data rates up to 50 kbit/s and is able to cover radio ranges in the order of kilometers \cite{Performance17}.
A typical metropolitan LoRaWAN deployment is constituted by multiple-gateways connected  to a common Network Server (NetServer).
The communication between end-devices and gateways is spread out on different frequency channels and data rates. LoRa uses up to 6 different programmable Spreading Factor (SF): 7, 8, 9, 10, 11, 12.
Furthermore, also the adopted bandwidth can be configured: $125\:kHz$, $250\:kHz$ and $500\:kHz$ (typically $125\:kHz$ for the $868$ ISM band.)
The selection of the data rate is a trade-off between communication range and message duration, given that communications with different SFs are often assumed not to interfere with each other. LoRa data rates range from $0.3\:kbp$s to $50\:kbps$. To maximize both battery life of the end-devices and overall network capacity, the LoRa network infrastructure can manage the data rate and RF output for each end-device individually by means of an Adaptive Data Rate (ADR) scheme. The relation between the nominal bit rate and the SF is given as:
$R=SF*\frac{CR}{2^{SF}/BW}$.\\
\indent In this context, several papers have analyzed the potentialities of LoraWAN systems in terms of scalability and performance \cite{Bor2016}\cite{Mitigating16} and have also highlighted the relevant limits \cite{Limits}.
Specific attention has been recently given to the power and spreading factor allocation in order to avoid the near-far problems by allocating distant users to different channels \cite{Power17} and to definition of mechanisms to configure the communication parameters of LoRa
networks in dense IoT scenario \cite{Adaptive18}.
In the paper \cite{Explora-AT} the EXPLoRa-Air Time (AT) has been defined. This solution playing with the LoRa modulation SF technique improves the throughput and data extraction rate performance compared whit the ADR \cite{LoRa}. The main result of EXPLoRa-AT is that there is the possibility, in the radio range of a single LoRa Gateway, to allocate different spreading factors to the transmitting end-devices, with the goal of assuring a similar time on air period to all of the them and inducting less collisions and an higher throughput.\\
\indent In this work we extend the use of the idea of balancing the Air-Time ($AT$) in a a multi-gateway scenario. To this aim we define an ADaptive Mitigation of the AIr-time pressure in lORa (AD MAIORA) for a multi-gateway scenario.\\
AD MAIORA is based on the following main considerations: i) there are multiple gateways that may receive the same message by the same end-device if a given SF is used; ii)  the use of the same SF by multiple end-devices sharing the same GWs, in the same coverage area, my overload the wireless medium causing collisions. Some SFs may result overloaded and the idea of letting the different SFs having a similar occupation (measured as Time-on-Air) goes in the direction of mitigating message collisions.\\
Since the final objective is to increase the Data Extraction Rate (DER), i.e., the amount of messages that arrive at least at one GW without corruption, AD MAIORA proposes an heuristic able to keep high the DER by reducing the pressure that some nodes put on some GWs by using a given SF.
%
%
\section{AD MAIORA - ADaptive Mitigation of the AIr-time pressure in lORa}
\label{sec:AD}
In a multi-gateway scenario, the assignment of SFs to nodes of the network has a different impact compared to scenarios with a single gateway.
In this case, in fact, we have to consider not only the impact that a node transmitting at a given SF has on a single GW, but also the impact it has on the other GWs of the network in its radio visibility. By changing the SF of a node we can give to it more radio visibility but this may also influence the overall network depending on the reciprocal distances of the node from all the gateways.\\
\indent The idea of AD MAIORA is to make a SF allocation that allows to distribute the load in a fair way not only on the different SFs but (as far as possible) on the different GWs.
AD MAIORA takes into account that we can assign higher SF values to nodes eligible for a lower SF (e.g. SF=8 for nodes at SF=7) while the opposite cannot be applied; furthermore, we start from a basic assignment that depends on sensitivity thresholds (the ones used in the ADR scheme \cite{Bor2016}).\\
To have an example, let's consider a scenario with two GWs and only two nodes both located in the overlapping area between GWs. Let's assume that both EDs are positioned in the  $SF8$-range of both GWs: with ADR we will have both nodes with the same SF $=8$, but we could also increase the SF of one ED in order to not create interferences with the other one. It is important to take into account that moving nodes to an high SF always causes an $AT$ increase since, for high SF values, the duration of the chirp increases too.
We want to level off the $AT$ values over all the SFs of each GW and we can assign higher SF to nodes (on a given gateway) only if this will not increase the maximum $AT$ for that gateway.

\subsection{$ADR_{MGW}$: ADR in multiple gateway scenarios}
In AD MAIORA we use an ADR version which is compatible with multiple gateway scenario, and we named it $ADR_{MGW}$. The algorithm is the same as the ADR for a single-gateway, except that, the SF assigned to a node is the lowest assuring at least the visibility to 1 gateway. In other words, if at $SF7$ a node is able to reach fewer gateways than at $SF8$, $ADR_{MGW}$ assigns $SF7$.

\subsection{AD MAIORA algorithm}
The AD MAIORA algorithm is based on the concept of \textit{pressure}, i.e. the weight in terms of $AT$ that quantifies the load on each GW for all the SFs.
Let us denote by $N$ the number of end devices (also denoted by nodes) in the considered multi gateway scenario and by $N_{GW}$ the number of gateways. Let is $GW$ the set of gateways in the considered area. Let us consider $\mathcal{RSSI}$, a $[N_{GW} \times N]$ matrix storing all the RSSI values measured in the network: the generic element $\mathcal{RSSI}[i,j]$ represents the RSSI value with which the $i-th$ GW receives the signal of $j-th$ ED. Given $N_{SF}$ possible SFs (7 in the LoRa legacy scheme) and $N_{BW}$ possible spectrum bands for the LoRA communications, we identify as $\mathcal{S}$ as a $[N_{SF} \times N_{BW}]$ matrix collecting the sensitivity thresholds: the generic element $\mathcal{S}[i,j]$ is the threshold for the $i-th$ SFs and the $j-th$ BW value. Let us denote $sf_{cost}$ as a $[1 \times N_{SF}]$ vector collecting the basic AT values for each SF i.e. $[1.0, 2.0, 3.56, 7.12, 14.23, 24.93]$. 
In LoRa, the AT (or packet transmission time) is a value depending on multiple parameters such as the SF $[7,\cdots, 12]$, the BW [125 $kHz$, 250 $kHz$, 500 $kHz$], the header and payloads lengths, the Code Rate (CR) and also depends on two flag variables i.e. DE (= 1 if low Data Rate Optimization is enabled) and H (= 0 when the header is enabled).\\
This can be expressed as $AT = T_{preamb} + T_{PL}$ that is the sum of the preamble time $T_{pream}$ and the payload time $T_{PL}$;  in order to compute these values we need to define the symbol time $T_{sym}=2^{SF}/BW$ depending on the SF and the BW. So, the duration of preamble is given by $T_{pream} = (n_{pream}+4.24)\cdot T_{sym}$ where $n_{pream}$ is the number of programmed preamble symbols. Instead, to compute $T_{PL}$ we need to evaluate how many symbols make up the packet payload and header as follows:
\begin{equation}
PLsym = 8 + \max[\ceil{ \dfrac{8PL-4SF+44-20H}{4(SF-2DE)} }\cdot(CR+4),0]
\end{equation}
\indent Once the \textit{AT} has been defined, we can define $sfpress$ as a [$N_{SF}\times N_{GW}$] matrix collecting the sum of the $AT$s for each SF and each GW: the generic element $sfpress[i,j]$ represents the sum of all $AT$s (in milliseconds) deriving by all nodes transmitting at the $i-th$ SF value on the $j-th$ GW.\\
\indent Our purpose is to reduce and balance the \textit{pressure} on the different GWs playing with the SFs and the different nodes visibilities to the GWs. To this aim we identify two variables: $sfmap$ and $nodetosf$. The former is the current mapping of SFs [$N \times N_{GW}$], the latter is the final mapping (after AD MAIORA) between nodes and the new SFs.
The $sfmap$ is initialized with $ADR_{MGW}$. The algorithm is characterized by multiple iterations during which the $sfmap$ variable is used in order to associate a temporary SF to a node. At the end of each run the value of the $i$-th element of $nodetosf$ is updated with the value computed in $sfmap$. The algorithm runs until all the changes are added to $nodetosf$. 
For each iteration, AD MAIORA performs three steps: first, calculates the $sfpress$; then, it finds the best node than can be moved to another SF in order to improve the overall performance; finally the algorithm identifies an optimal SF value and, if there is "free Air Time" left, both node and new SF are added to $nodetosf$.
In the following subsections \ref{subsec:choosingnode} and \ref{subsec:choosingsf} we will explain in detail how the best node and the best SF are chosen.

\subsection{Choosing the best node}
\label{subsec:choosingnode}
In order to find the best candidate node for a movement from a SF to another SF, we derive from the pressure table ($sfpress$) the most stressed gateway ($worstGW$) and the most overloaded SF for that gateway ($wSF$). Then we can build a set of nodes, named $stressingNodes$, composed by nodes transmitting with $SF = wSF$, in the coverage area of the GW $worstGW$.
Let us introduce a weight $W_n$ for each node $n$ belonging to $stressingNodes$ that represents the benefit obtained (in terms of AT) if we set a greater SF to the node $n$. 
Since we use $ADR_{MGW}$ to assign the initial SFs values, we can define $SF^*$ as a set comprising the overall spreading factor values higher than $wSF$, and we can assume $sf^*$ as an element of the set above.\\
\indent We compute for each $i$-th GW and node $n\in stressingNodes$, $\Delta_{gw_i, sf^*}$ for all values in $SF^*$ as the difference between the maximum $\lambda_{gw_i}$ 
(defined as in Eq. \ref{lambda}) and the sum of ATs caused by nodes with $sf^*$ to GW $gw_i$\footnote{The $i$-th gateway is considered for node $n$ only if $n$ is visible to the gateway and the $\mathcal{RSSI}[gw_i, n] \geq \mathcal{S}[sf^*, bw]$} as in Eq. \ref{diff}:
\begin{equation}\label{lambda}
\lambda_{gw_i} = Max(\{sfpress[sf, gw_i] | \forall sf\in SF\})
\end{equation}
\begin{equation}\label{diff}
\begin{split}
\Delta_{gw_i}^{(n)}& = \{\lambda_{gw_i} - sfpress[sf^*, gw_i]\\
  \hspace{2 mm}&   |\forall sf^* \in SF^*, sf^*>wSF\}
\end{split}
\end{equation}
Hence, for all GW, we compute the minimum $Min(\Delta^{(n)}_{gw_i})$; among the possible $\Delta_{gw_i}^{(n)}$ values we take the smallest for each GW because, for that node $n$, it allows to change SF for that node by keeping low the increment of total AT.\\
We add then this value at the set $\Delta$ collecting all minimum values for all GWs; finally, for each node $n$, $W_n$ is:  
\begin{equation}
W_n = \sum\limits_{i=1}^{N_{GW}} Min(\Delta^{(n)}_{gw_i})
\end{equation}
The "best node" is the one having the highest $W_n$: among all possible $W_n$ values belonging to stressed nodes $n$, we take the highest because we would have more degrees of freedom, i.e. each node can communicate with more GWs.
Essentially, we are looking for the "free space" that each (reachable) gateway has, for every SF and for every node (reachable GWs and SFs sets change due to different locations of nodes and overlapping regions of GWs).
\begin{algorithm}[t]
	\small
	\caption{Choose the best node}
	\label{algo:multigw_algorithm_p1}
	\begin{algorithmic}[1]
		\Function{\texttt{$bestNode$}}{$\mathcal{RSSI}, sfmap, sfpress, \mathcal{S}, GW, N, bw$}
    \State $wSF, worstGW = Max(sfpress)$
    \State $stressingNodes = \{n | \forall n \in N, sfmap[worstGW, n] = wSF\}$
    \State $candidates = \{\}$
    \For{$n \in stressingNodes$}
      \State $\Delta = \{\}$
      \For{$gw \in GW$}
        \State $\Delta^{(n)}_{gw_i} = \{\}$
        \For{$sf^* \in [7, 8, 9, 10, 12, 11] \textbf{ and } sf^* > wSF$}
          \If{$\mathcal{RSSI}[gw_i, n] \geq \mathcal{S}[sf^*, bw]$}
            \State $\lambda_{gw_i} = Max(sfpress[\cdot, gw_i])$
            \If{$\lambda_{gw_i} > sfpress[sf^*, gw_i]$}
              \State $\Delta^{(n)}_{gw_i} = \Delta^{(n)}_{gw_i} \bigcup \{\lambda_{gw_i} - sfpress[sf^*, gw_i]\}$
            \EndIf
          \EndIf
        \EndFor
        \State $\Delta = \Delta \bigcup \{Min(\Delta^{(n)}_{gw_i})\}$
      \EndFor
      \State $W_n = \sum{\Delta}$
      \State $candidates = candidates \bigcup \{(n, W_n\}$
    \EndFor
    \State \Return $pickNodeWithGreaterW_n(candidates), wSF$
		\EndFunction
	\end{algorithmic}
\end{algorithm}

\subsection{Find the best spreading factor}
\label{subsec:choosingsf}
Once the best node has been found, we need to evaluate which SF value could be the best choice. In order to find the optimal SF, we need to evaluate how changing the SF impacts on the network (the calculus is similar to the one considered in \ref{subsec:choosingnode}, except that now we are subtracting also $sf_{cost}[sf^*]$ as in Eq. \ref{deltaprime}).
Hence, the heuristic calculates $\Delta^\prime_{sf^*}$ collecting all the differences between $\lambda_{gw_i}$, the pressure on the current GW at $sf^*$ and the cost of that node at $sf^*$ in terms of Air Time; this is stored in the set $\Delta^\prime_{sf^*}$ computed as in Eq. \ref{deltaprime} for all $SF^*$ values of the node selected in \ref{subsec:choosingnode}.
The best SF is the one that allows to get the highest $Min(\Delta^\prime_{sf^*})$ as in \eqref{nextat}).
\begin{equation}\label{deltaprime}
\begin{split}
	\Delta^\prime_{sf^*}&=\{\lambda_{gw_i}-sfpress[sf^*, gw_i]-sfcost[sf^*]\\& | \forall gw_i \in GW\ \}
\end{split}
\end{equation}
\begin{equation}\label{nextat}
	nextAT = Max(\{Min(\Delta^\prime_{sf^*}) | \forall sf^* \in SF^*\})
\end{equation}

Essentially, we are looking for the spreading factor with the maximum free space among all considered gateways.

\begin{algorithm}[t]
	\small
	\caption{Find the best spreading factor}
	\label{algo:multigw_algorithm_p2}
	\begin{algorithmic}[1]
		\Function{\texttt{$bestSF$}}{$n, wSF, \mathcal{RSSI}, sfpress,
		sfcost, \mathcal{S}, GW, bw$}
    \State $nextSF = 0, nextAT = 0$
    \For{$sf^* \in [7, 8, 9, 10, 12, 11] \textbf{ and } sf^* > wSF$}
      \State $\Delta^\prime_{sf^*}= \{\}$
      \For{$gw \in GW$}
        \If{$\mathcal{RSSI}[gw_i, n] \geq \mathcal{S}[sf^*, bw]$}
          \State $\lambda_{gw_i} = Max(sfpress[\cdot, gw_i])$
          \If{$\lambda_{gw_i} > sfpress[sf^*, gw_i]$}
            \State $\Delta^\prime_{sf^*} = \Delta^\prime_{sf^*} \bigcup \{\lambda_{gw_i} - sfpress[sf^*, gw_i] - sfcost[sf^*]\}$
          \EndIf
        \EndIf
      \EndFor
      \If{$nextAT < Min(\Delta^\prime_{sf^*})$}
        \State $nextAT = Min(\Delta^\prime_{sf^*})$
        \State $nextSF = sf^*$
      \EndIf
    \EndFor
		\State \Return $nextAT, nextSF$
		\EndFunction
	\end{algorithmic}
\end{algorithm}

\section{Simulation model}
\label{sec:sim}
\begin{figure*}[t]
	\center
	\subfigure[Balanced topology]{\includegraphics[width=0.67\columnwidth]{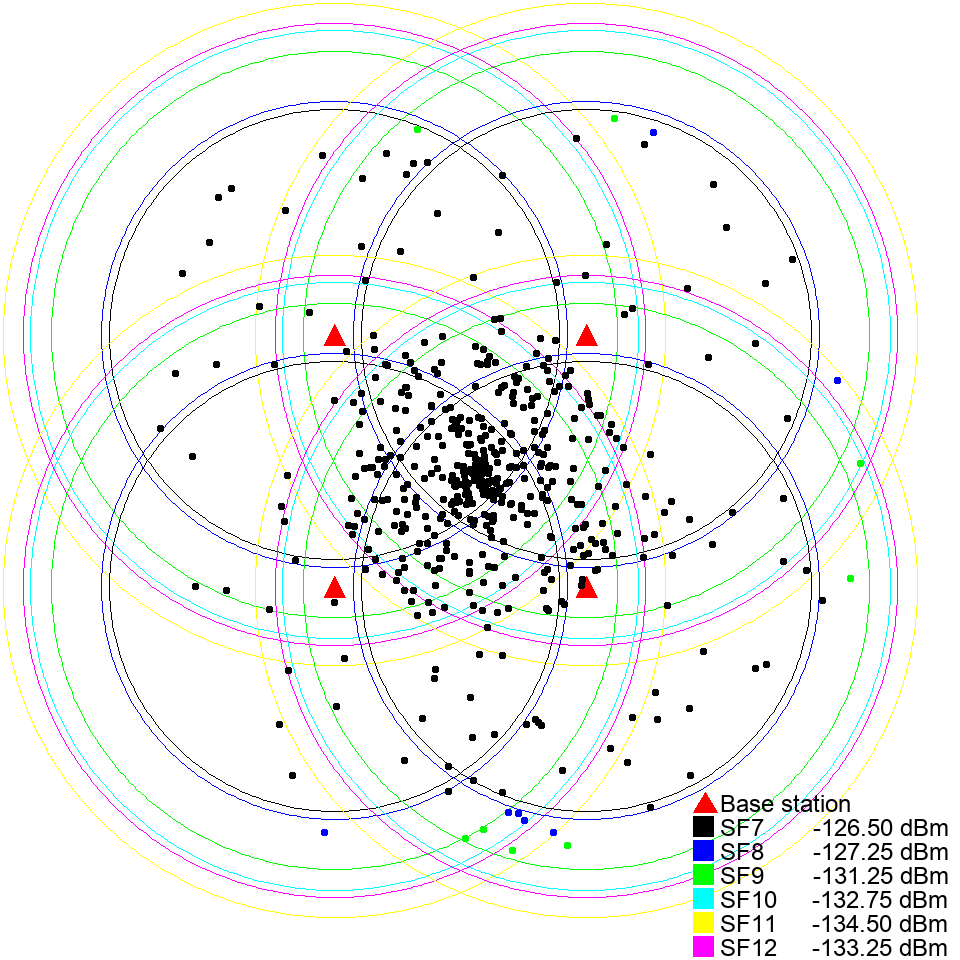}\label{fig:balanced}}\hspace{10 mm}
	\subfigure[Unbalanced topology] {\includegraphics[width=0.67\columnwidth]{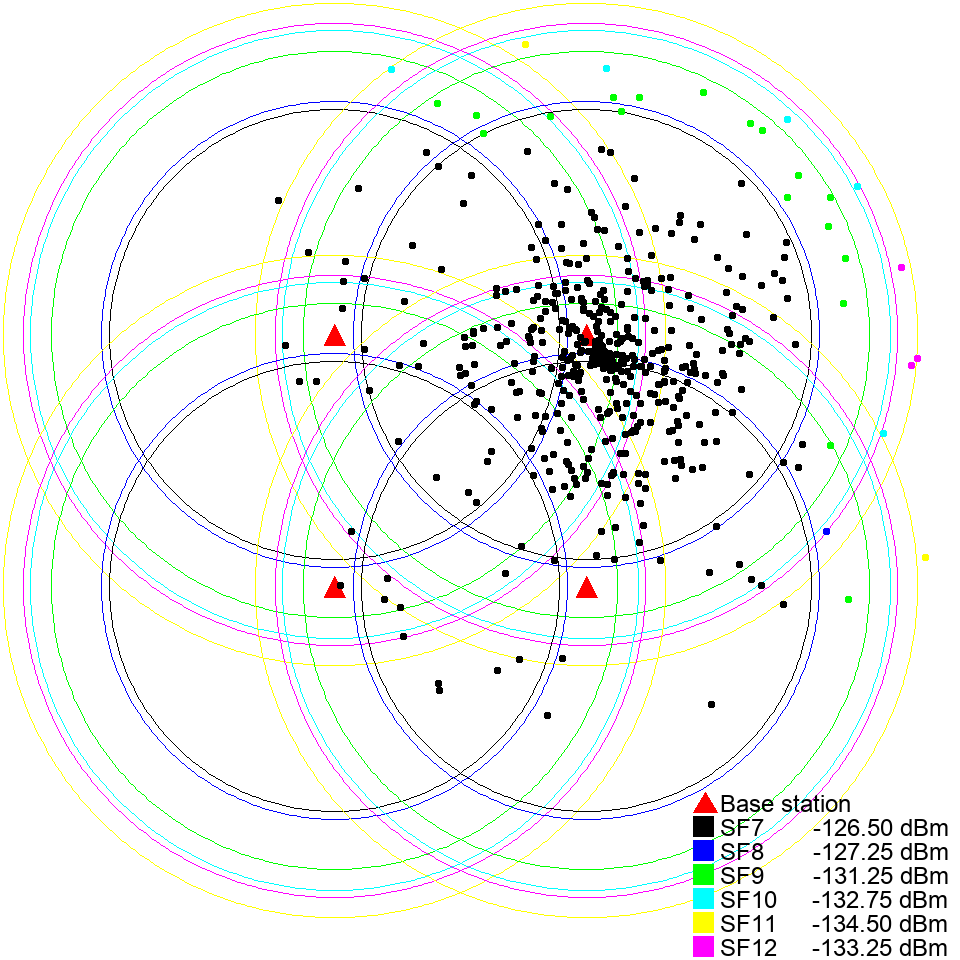}\label{fig:unbalanced}}\\
	\subfigure[ADR connections]
	{\includegraphics[width=0.66\columnwidth]{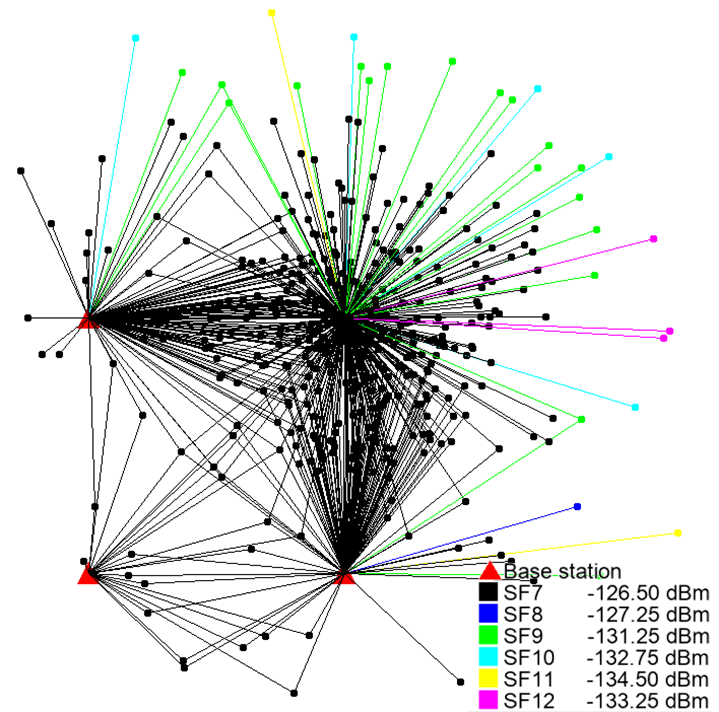}\label{lorasim_pre_spider}} \hspace{10 mm}
	\subfigure[AD MAIORA Connections]
	{\includegraphics[width=0.66\columnwidth]{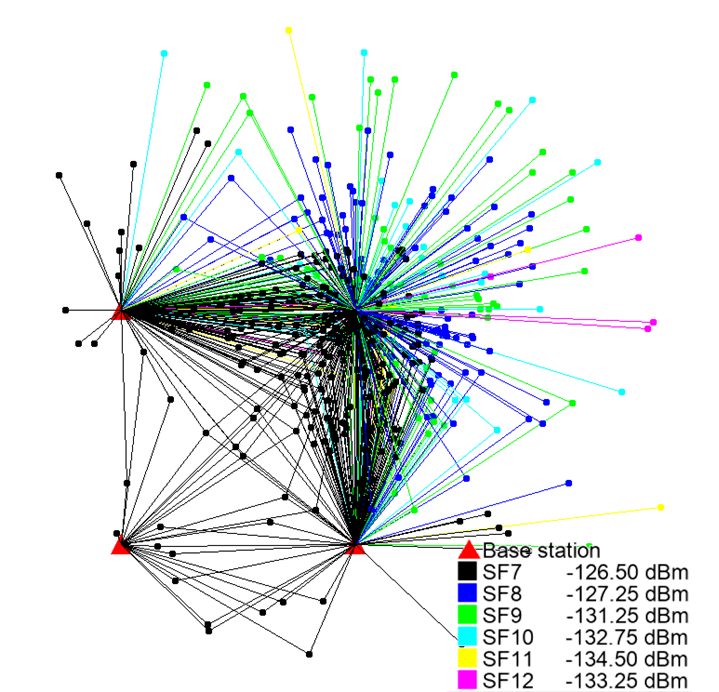}
		\label{lorasim_spider}}
	\caption{Nodes topology and ADR allocation of SFs (different colors of the nodes) for 500 EDs: (a) Balanced (BAL) and (b) Unbalanced (UNBAL); SF allocations (represented by different colors) for all possible connections in the unbalanced scenario  (c) ADR connections and (d) AD MAIORA connections}
	\label{fig:position_nodes}
\end{figure*}

The performance evaluation of AD MAIORA has been carried out by extending LoRasim simulator \cite{Bor2016}. We considered different scenarios where $N$ End-Devices (EDs) are randomly distributed in a bi-dimensional space around one or more GWs. We analyzed two different topologies for the node's location as shown in Figure \ref{fig:position_nodes}: a balanced one in Figure \ref{fig:balanced} where the 60\% of nodes is located in a central area between the gateways, and an unbalanced one, in Figure \ref{fig:unbalanced}, where 60\% of nodes are located around a specific gateway in a range of 50 meters.  All simulation results are represented with their 95\% confidence interval. The nodes work using the communication transmission parameters reported in Table \ref{tab:Simul}.

Regarding the duty cycle, in Europe its values are established by ETSI EN 300.220 standard \cite{etsi}, which defines different values for different sub-bands: in particular, one of those sub-bands ($G3: 869.4 - 869.65 MHz$) has a duty cycle of 10\%. This means that considering a message period equal to 10 seconds, at the maximum air time value (quantified at about 1 second for SF12) there are 9 seconds of waiting time before a new retransmission \cite{Limits} \cite{Alliance}.
Since in the simulations we considered a message period equal to 10 seconds, we are exactly in the case where the duty cycle is at 10\% (not higher).

The goal of our simulations is to prove the effectiveness of the algorithms at a high pressure level: using the $G3$ sub-band
we get a simulation where the devices transmit much more frequently (so with greater probability of collision).
The stress level in our simulations is therefore the maximum possible, while remaining within the limits imposed by LoRaWAN and the standard for the ISM band.
\begin{table}[t]
	\caption{Simulation parameters.}
	\label{tab:Simul}
	\centering
	\begin{tabular}[]{lc}
		\toprule
		\textbf{Parameter} & \textbf{Value} \\
		\midrule
		Carrier Frequency (MHz)& $869.5$  \\
		Bandwidth (kHz)& $125$  \\
		Code Rate (CR) & 4/5 \\
		Duty cycle [\%]& [0.1-10]\\
		Message size [bytes]& 20\\
		Message Period - MP [sec]& [10-900]\\
		Number of gateways & 1-2-4-8\\
		Number of nodes& 50-100-250-500-1000\\
		Path loss & Eq. (3) of \cite{Bor2016} with $\overline{L_{pl}}(d_{0})=127.41\;dB$\\
		&$d_{0}=40\;m$ $\gamma=2.08$, $\sigma^{2}=0$\\
		\bottomrule
	\end{tabular}
\end{table}
Furthermore, in our simulations it has been proposed an allocation algorithm called "probabilistic ADR" that distributes the SFs following a probability distribution inversely proportional to the air time. In this way, considering only the number of nodes for each SF value (and not their position relatively to GWs), we will get a term of comparison of AD MAIORA against a simpler approach. 

\section{Simulation results}
\label{sec:result}
\begin{figure*}[t]
	\centering
	\subfigure[DER]
	{\includegraphics[width=0.67\columnwidth]{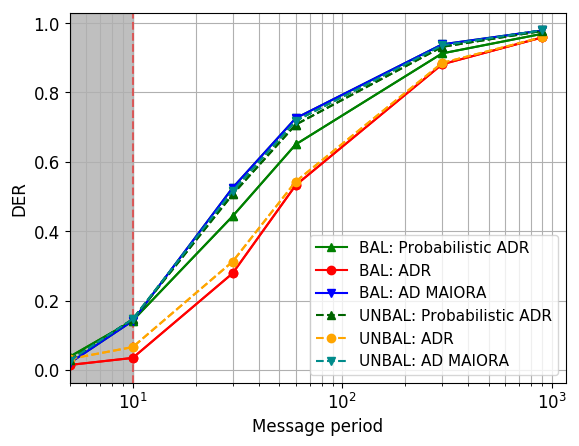}
		\label{DER_vs_MP_t67}}\hspace{10 mm}
	\subfigure[Throughput]
	{\includegraphics[width=0.67\columnwidth]{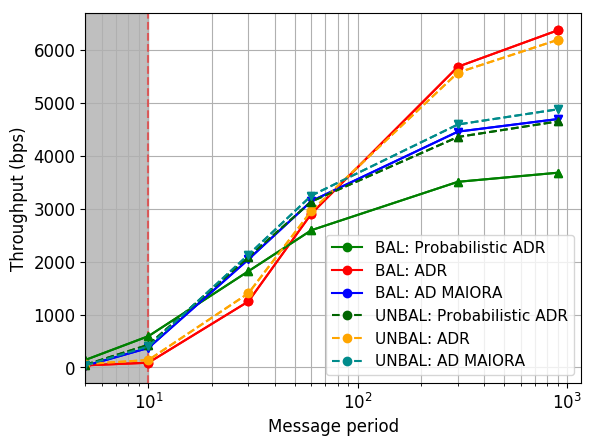}
		\label{th_vs_MP_t67}}
	\caption{DER and throughput as a function of the Message Period: case of 500 EDs, 4 GWs, scenarios \ref{fig:balanced} and \ref{fig:unbalanced}}
	\label{fig:performance_vs_MP_t67}
\end{figure*}

\begin{figure*}[t]
	\centering
	\subfigure[DER]
	{\includegraphics[width=0.67\columnwidth]{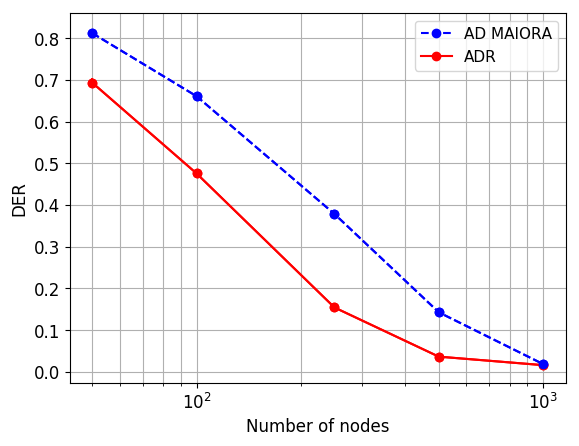}
		\label{DER_vs_n_t5}}\hspace{10 mm}
	\subfigure[Throughput]
	{\includegraphics[width=0.67\columnwidth]{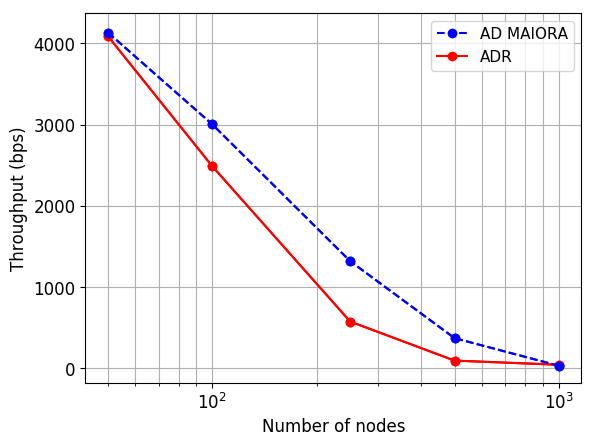}
		\label{th_vs_n_t5}}
	\caption{DER and throughput as a function of the number of nodes: case of 4 GWs, MP=10 sec, scenario \ref{fig:balanced}}
	\label{fig:performance_vs_n_t5}
\end{figure*}

\begin{figure*}[t]
	\centering
	\subfigure[DER]
	{\includegraphics[width=0.67\columnwidth]{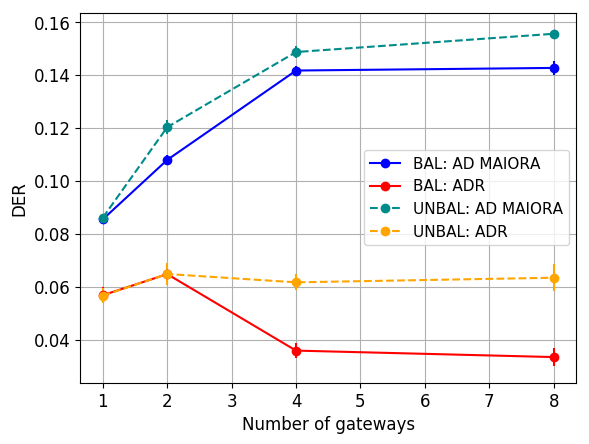}
		\label{DER_vs_GW_t12}}\hspace{10 mm}
	\subfigure[Throughput]
	{\includegraphics[width=0.67\columnwidth]{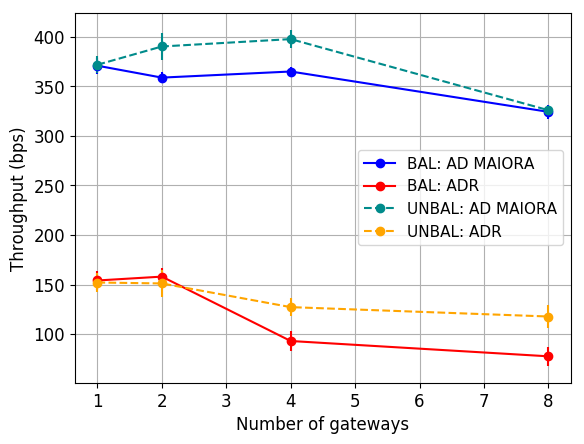}
		\label{th_vs_GW_t12}}
	\caption{DER and throughput as a function of the number of GWs; case of 500 EDs, MP=10 sec, scenarios of Fig.~\ref{fig:balanced} and Fig.~\ref{fig:unbalanced}}
	\label{fig:performance_vs_GW_t12}
\end{figure*}

\begin{figure}[t]
	\centering
	\includegraphics[width=0.67\columnwidth]{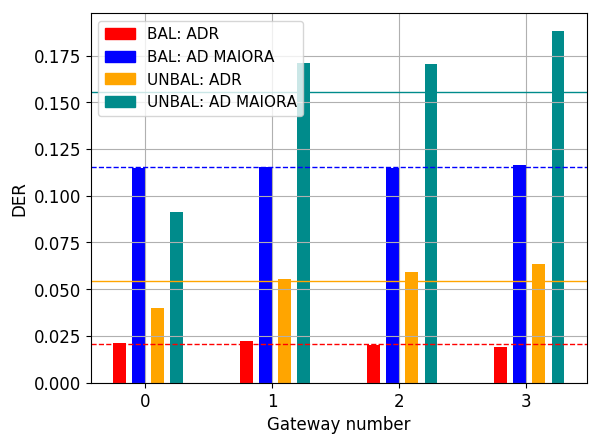}
	\caption{Partial DER for each GW; case of 4 GWs, 500 EDs, MP=10s, scenarios of Fig.~\ref{fig:balanced} and Fig.~\ref{fig:unbalanced}}
	\label{fig:T89_DER}
\end{figure}

\begin{figure}[t]
	\centering
	\subfigure[ADR]
	{\includegraphics[width=0.67\columnwidth]{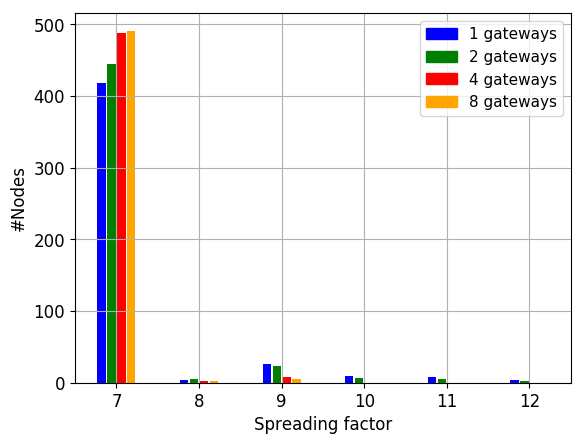}
		\label{SF_vs_nodes_adr}}
	\subfigure[AD MAIORA]
	{
		\includegraphics[width=0.67\columnwidth]{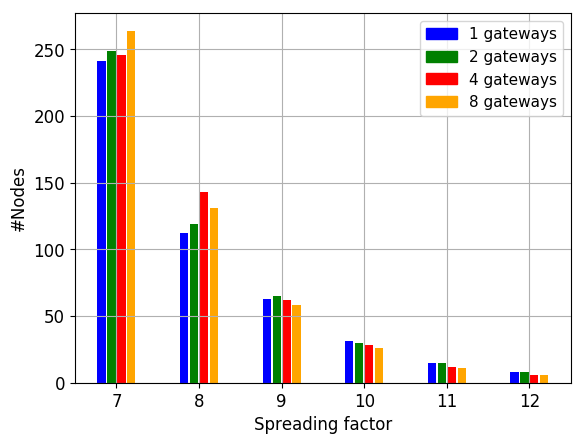}
			\label{SF_vs_nodes_admaiora}}
	\caption{Number of EDs as a function of SF value for a) ADR allocation and b) AD MAIORA allocation; case of 500 EDs, MP = 10s, scenario of Fig.~\ref{fig:balanced}}
	\label{fig:SF_vs_nodes}
\end{figure}

As a first evaluation, in Figure \ref{fig:performance_vs_MP_t67} we plot the DER and throughput as a function of the message period with six different lines, each one representing different scenarios and allocations.
The scenario is characterized by 4 GWs and 500 EDs, located according to the two topologies considered above. Each ED forwards a message with a decreasing rate. We note that in both scenarios the ADR attains the highest throughput only for high message periods (low message rates) while AD MAIORA presents improvements at high network loads (i.e. high message rates). The figure shows a gray area representing the values of message period violating the possible duty cycle. In order to  clearly plot the different scenarios, we used continuous lines for balanced scenarios and dashed lines for unbalanced onces. In case of balanced scenarios, with AD MAIORA there is a DER improvement, especially in case of lower message periods; in the other case we have for probabilistic ADR both good DER and throughput because the assignment made with AD MAIORA (when the 60\% of node are positioned around a single GW) is similar to EXPLoRa-AT: this is due to the fact that in both cases the number of EDs with the same SF value is inversely proportional to the air time.
Analyzing the plot in Figure \ref{th_vs_MP_t67} we can derive that, for a message period greater than 100 seconds, the ADR has a good throughput so it would not be possible to bring improvements. On the contrary, we can improve the performance in case of highly stressed scenarios, e.g. when nodes transmit with a message period equal to 10 seconds.\\
\indent In Figure \ref{fig:performance_vs_n_t5} we evaluate the performance in terms of DER and throughput as a function of the number of nodes. In this case, each device transmits with a message period equal to 10 seconds, and nodes are placed as in Figure \ref{fig:balanced}.
As expected, by increasing the number of nodes, both the DER and the throughput decrease following a monotonous trend, but it is possible to verify that with AD MAIORA the performance is better than ADR of a nearly constant value, until the devices do not exceed a certain number. In fact, we can observe that the performance drops significantly when the gateways are subjected to a great pressure, that is if the nodes start to become more than 250. For this reason, in the following experiments we analyze scenarios with a number of nodes equal to 500, in order to evaluate the performance of AD MAIORA in critical contexts highlighting its ability to make the best use of the GWs present in the scenario.\\
\indent We also plotted the performance according to the GWs number considering a scenario with fixed nodes distributed as the above topologies. In case of single-GW, 500 EDs are distributed in a circular range of 50 meters around the GW.
The DER and throughput behaviors are shown in Figure \ref{fig:performance_vs_GW_t12} for two different spreading factor allocations: i) $ADR_{MGV}$ and ii) AD MAIORA, where both are considered for the balanced (Fig.~\ref{fig:balanced}) (continuous lines) and unbalanced (Fig.~\ref{fig:unbalanced}) (dashed lines). The message period is set to 10 seconds. We notice that AD MAIORA attains a good performance gain as the number of GWs increases.
In particular, we can evaluate that ADR in both scenarios (red and orange lines) has an unusual trend between values 1 and 4 in abscissa, so passing from a single-gateway scenario to a 4-gateways one.
This irregular trend can be explained by observing Fig.~\ref{fig:SF_vs_nodes} which shows how many nodes have the same SF value when the number of gateways increases in the topology: we can infer that the gateways increment leads to a greater radio coverage, on the other hand we will have a greater number of nodes having the same SF value due to the overlapping areas. Indeed, we can notice in Fig.~\ref{SF_vs_nodes_adr} that in case of 8 GWs and ADR allocation, almost all nodes have the same $SF=7$ value, instead Fig.~\ref{SF_vs_nodes_admaiora} shows that the number of nodes assigned for each SF value is inversely proportional to the respective air time, like EXPLoRa-AT in \cite{Explora-AT}.
Furthermore, in the same figure, we show the performance evaluated in case of Fig.~\ref{fig:unbalanced} plotted with dashed lines. We can notice a more continuous trend of the ADR curve (yellow) and a slightly higher performance compared to the balanced topology case; we can explain this behavior by considering the partial DER for each gateway: as shown in Fig.~\ref{fig:T89_DER} in case of balanced topology (red and blue bars), each gateway has approximately the same DER value because the majority of nodes are in the overlap central areas and so, for instance, a node having $SF=9$ located in the middle area overloads the channel of all gateways. Instead, in case of unbalanced topology (orange and green bars), there is also an unbalanced DER for each gateway; in particular, there is a gateway with a very low DER corresponding to the overloaded one, but others have a good DER due to a low traffic load.

\section{Conclusions}
\label{sec:conc}
In the IoT market, the Low Power, Low Range technologies (LoRa) are emerging. To fully exploit the LoRa potentials it is critical to support hundreds of devices also in stressed configurations (e.g. multiple gateways interconnected to nodes transmitting with a 10\% of duty cycle and a high message rate). To this aim, in this paper, we presented AD MAIORA, an algorithm that adaptively allocates the spreading factor to nodes in the radio visibility of multiple gateway with the aim of reducing the pressure on the gateways of nodes using the same SFs. We showed, via simulations, that AD MAIORA, by suitably allocating the SFs attains a noticeable gain with respect to the classic ADR approach. This result is particularly evident in case of very loaded and unbalanced scenarios.

\bibliography{biblio}
\bibliographystyle{IEEEtran}

\end{document}